# A charging method for electric vehicle using multi battery series mode

Hong Liu[*], Yuan Ren, Liangxiao Ye
School of Electrical Information, Shanghai Dianji University, Shanghai 200240, China
[*]Corresponding author: liuhong@sdju.edu.cn

*Abstract*—A charging method for electric vehicle using multi battery series mode which consisted of the following steps was introduced: the battery series is firstly charged at a constant power with a charging current of I1. When the terminal voltage over the battery series has reached the 1st threshold voltage, the charging current will reduce to I2 and the power remains constant. When the terminal voltage over the battery series has reached the 2nd threshold voltage, the charging current will reduce to I3 and the power remains constant (the 2nd threshold voltage is larger than 1st threshold voltage). When the terminal voltage over the battery series has reached the 3rd threshold voltage, which is the rated voltage of the battery series, the charging process is completed (3rd threshold voltage is larger than 2nd threshold voltage). Contents in this paper provide a charging method for electric vehicles using multi battery series mode with high charging efficiency and security level.

*Keywords—charging device; battery mode; threshold voltage*

## I. INTRODUCTION

Compared with traditional motor vehicles powered by fuel, electric vehicle is advantageous in energy saving, emission reduction and environmental protection. As a result, it is becoming more and more important with more applications in transportation [1-8]. Up to now, most of the charging services of electric vehicles in social communities are realized by using appointed battery replacement managed by the State Grid. However, during the last couple of years, large scale electric vehicle rental services and bidding projects of charging station construction have been started in metropolitans such as Shanghai and Beijing. Besides, self-service electric vehicle charging posts have been constructed in a large scale around residential communities, public parking lots and business districts in large and medium-sized cities all over China. Development in all these aspects indicates that the electric vehicle charging services are becoming multi-functional, multi-zone and intelligent on the basis of traditional battery replacement service [9-15]. Therefore, the popularization and application of electric vehicle charging service is inevitably related to the construction and completion of the relevant charging infrastructures, which imposes higher requirements on the use and function of the electric vehicle charging facilities [16-21]. Related literatures showed that users would like the vehicles to be more intelligent, allowing them to select among functions modules such as the charging duration, charging speed mode and scheduled charging, which facilitates the use of vehicles under different conditions [22-24].

The selection of charging mode has equal importance for battery distribution management and charging security [25]. For example, some users rely on direct current (DC) source to finish vehicle charging process within a short time. However, the relatively high charging current from DC source will reduce the lifespan of batteries to a large extent. Hence, this charging method can only serve as an emergent backup plan. Researchers both at home and abroad found that there generally exist problems such as low energy conversion efficiency, instauration of battery when fully charged, battery heating and deformation, among which the low charging efficiency has the most obvious effect [26-28]. It reduces the mileage of electric vehicles, causes frequent charging in a short time and shortens the lifespan of batteries. In order to solve the above problem, Li et al. have done deep investigations into the charging management mode of the batteries in electric vehicles and the design of control system which balances the load current was found to be the key factor [29]. Thanks to their charging model proposed in their latest patent [29], based on his method, we will further present in this paper the design of a corresponding control system or facility for battery charge control, which will be greatly important in high current charging with DC source. In addition, the control design should provide a sustainable and highly secured operation environment with high energy utilization efficiency for electric vehicle users.

Therefore, in order to guarantee the lifespan, security and endurance of the batteries in electric vehicles and improve the energy conversion efficiency during battery charge process, it is imperative to establish a charging method for multi battery series which is complete, secure and easy to use. The investigation of this method is of great importance to reduce electric energy loss, energy saving, low carbon emission and environment protection. As a consequence, our paper intents to give the detailed description for the proposed and further carry out experimental results to support our discoveries.

## II. THE FUNDAMENTAL BASIS FOR CHARGING MULTI BATTERY SERIES

In order to overcome the shortcoming of the current electric vehicle charging method using multi battery series mode, such as low charging efficiency and low level security, this paper proposes a charging method with high charging efficiency and high level security.

The technical procedure for solving the problem mentioned should be specific, in general. According to the newest model proposed in Li's patent [29], the charging method for electric vehicles using multi battery series generally consists of the following steps: firstly, the battery series is charged at a constant power with a charging current of I1. Afterwards, when the terminal voltage over the battery series reaches the 1st threshold voltage, the charging current will reduce to I2 and the power remains as a constant. When the terminal voltage over the battery series reaches the 2nd threshold voltage, the charging current will reduce to I3 and the power remains as another constant (the 2nd threshold voltage is larger than 1st threshold voltage). When the terminal voltage over the battery series reaches the 3rd threshold voltage, which is the rated voltage of the battery series, the charging process is completed (the 3rd threshold voltage is larger than the 2nd threshold voltage).

One step further, the charging controller operated in dynamic control mode is able to carry out real-time control of the input current and terminal voltage of the battery series. The proposed method is advantageous in that the whole charging process is divided into several individual steps according to the relation between charging voltage and input current, resulting in high charging efficiency. Besides, the security of this charging method can be verified using real-time simulations.

### III. THE ADVANTAGES OF THE PROPOSED METHOD WITH HIGH CHARGING EFFICIENCY AND LEVEL SECURITY

According to Figure 1 to 4, the proposed charging method for electric vehicles using multi battery series consists of following elements: the battery series is firstly charged at a constant power with a charging current of I1. When the terminal voltage over the battery series reaches the 1st threshold voltage, the charging current will reduce to I2 and the power remains as a constant. When the terminal voltage over the battery series reaches the 2nd threshold voltage, the charging current will reduce to I3 and the power remains as another constant (the 2nd threshold voltage is larger than the 1st threshold voltage). When the terminal voltage over the battery series reaches the 3rd threshold voltage, which is the rated voltage of the battery series, the charging process is completed (the 3rd threshold voltage is larger than the 2nd threshold voltage).

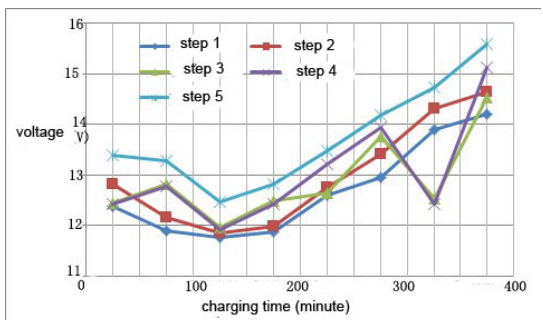

Fig. 1 The battery terminal voltage varying with time when the external current is input into the battery series in the electric vehicle.

One step further, the charging controller operated in dynamic control mode is able to carry out real-time control of the input current and terminal voltage of the battery series.

Figure 2 is the schematic diagram of the charging system of the multi battery series in the electric vehicle. As shown in Figure 2, the charging facility consists of battery charge controller (10), battery charger (20) and battery series (30). The main function of battery charge controller is to inspect the operation status of the battery series and output control signal to adjust the charging mode of battery series (30). In this paper, we adopt the charging method devised by Li et al. from their patent [29] where the main function of battery charger is to convert the external AC input into DC output. Besides, when the switch S is "ON", a closed loop of the DC current which is output by the battery charger can be formed, realizing the charging of battery series.

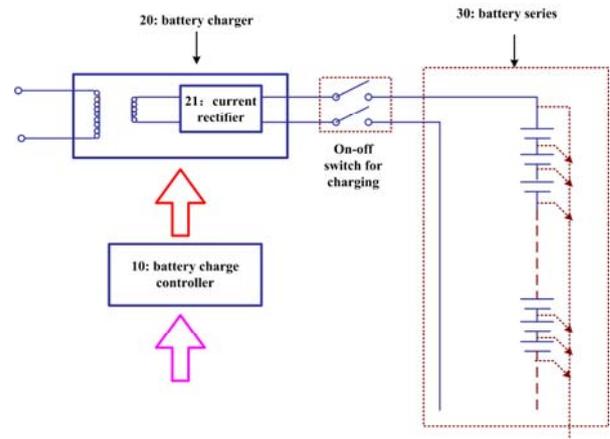

Fig. 2 The schematic diagram of the charging system of the multi battery series in the electric vehicle.

Subsequently, with the help of Figure 2 to 4, we will give detailed implementation of the multi battery series charging method for electric vehicles proposed in this section. Figure 3 is the schematic flow chart of charging steps of the multi battery series in the electric vehicle. Figure 4 is the battery terminal voltage varying with time and can be obtained from onsite experiments. When an electric vehicle breaks down due to battery energy depletion, the user would connect the charging plug to AC source and turn on the switch S to form a closed loop, in order to charge the battery series. During the process, the charging process s20 (as shown in Figure 3) will require the battery series charge controller (10) to identify the initial SOC of the battery series, and send control signal to battery charger (20) to instruct the system to charge the battery series (30).

Similarly, with the help of Figure 2 to 4, we will give detailed implementation of the multi battery series charging method for electric vehicles, which was proposed in this paper. Figure 3 is the schematic flow chart of charging steps of the multi battery series in the electric vehicle. Figure 4 is the battery terminal voltage varying with time which can be obtained from onsite experiments. When an electric vehicle breaks down due to battery energy depletion, the user could connect the charging plug to AC source and turn on switch S to form a closed loop in order to charge the battery series. During the process, the charging process s20 (as shown in Figure 3) will require the battery series charge controller (10) to identify the initial SOC of the battery series, and send control signal to battery charger (20) to instruct the system to charge the battery series (30).

Particularly, the 3 phase AC current input by user can be converted into DC output using the current rectifier (21) in the battery charger (20). During the charging process s10, the charging power maintains at 5.4 kW by the output current. Meanwhile, the charging process s20 requires real-time monitoring of the terminal voltage of the battery series by the battery charging controller (10).

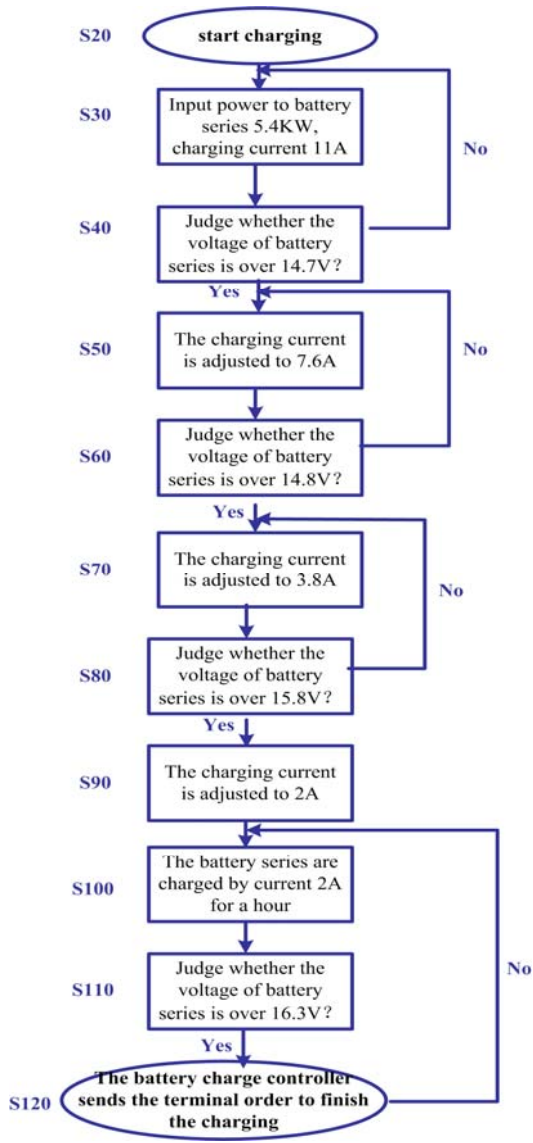

Fig. 3 The schematic flow chart of charging steps of the multi battery series in the electric vehicle.

When the terminal voltage of each sub-unit of the battery series reaches 14.7 V or the voltage of any sub-unit reaches 15.0 V (denoted as the charging process s30), the battery charging controller (10) will send charging command to battery charger (20), requiring the battery charger to charge the battery series with 11 A DC. The change of the terminal voltage of the battery series during this charging step is indicated by the waveform in Figure 4. When the charge process exceeds s40, the terminal voltage of the battery series will reach 14.8 V or any sub-unit of the battery series will reach 15.1 V, command will be sent again from battery charge controller (10) to battery charger (20). Battery charger will then reduce the charging current to 7.6 A, which is shown by the charging process s50 in

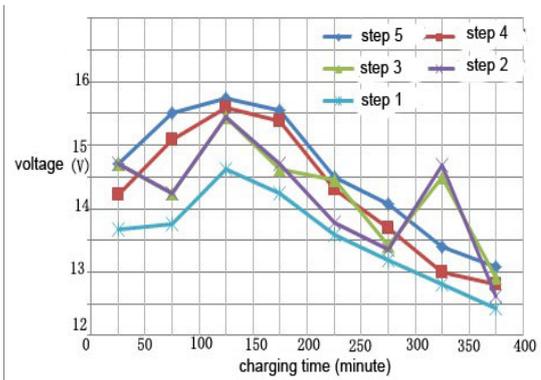

Figure 3.

Fig. 4 The battery terminal voltage varying with time, the corresponding results are obtained from onsite experiments.

When the terminal voltage of each sub-unit of the battery series reaches 14.7 V or the voltage of any sub-unit reaches 15.0 V (denoted as the charging process s30), the battery charging controller (10) will send charging command to battery charger (20), requiring the battery charger to charge the battery series with 11 A DC. The change of the terminal voltage of the battery series during this charging step is indicated by the waveform in Figure 4. When the charge process exceeds s40, the terminal voltage of the battery series will reach 14.8 V or any sub-unit of the battery series will reach 15.1 V, command will be sent again from battery charge controller (10) to battery charger (20). Battery charger will then reduce the charging current to 7.6 A, which is shown by the charging process s50 in Figure 3.

Particularly, the above charging process of the battery series would repeat for several times [29], which can effectively save energy losses, hence we use the model of Li's patent in our design procedures. To verify Li's such theorem, we show the comparison results in Figure 3, when the terminal voltage of the battery series reaches 14.8 V or the voltage of its sub-unit reaches 15.1 V for the first time, the battery charger will reduce the charging current to 7.6 A DC, which is the charging process s60 in Figure 3. After s60, the charge controller (10) will again send command to battery charger

(20), which is the charging process s80 in Figure 3. The battery charger will further reduce the current to 3.8 A or 2 A to charge the battery series. These two charging processes, s70 and s90 s90 is completed, if the terminal voltage of the battery series (30) further increases to 15.8 V or its sub-unit voltage increases to 16.1 V, the charge controller (10) will send new command to battery charger. The battery charger will then charge the battery series with a constant input voltage of 15.8 V, which is shown by s100 and s110 in Figure 3.

After charging process s100 has been executed for one hour, if the charging current of battery series (30) is less than 0.2 A or the sub-unit voltage is over 16.3 V, the charge controller (10) will send termination command to battery charger, which will stop inputting charging voltage into battery series and stop the whole charging process, as indicated by the theorem in [29].

This method proposes a novel approach for charging battery series in electric vehicles based on multi-steps. It is advantageous in reducing the effect of mutual interference between battery series sub-units to the maximum extent and achieving a battery energy usage efficiency of 80%. Specifically, the whole process of obtaining external electric energy for charging battery series is completely by implementing the charge controller. The charge controller carries out real-time control of input current and terminal voltage of the battery series in electric vehicles via dynamic control mode, which avoids damage to the batteries due to over-charging and high-speed charging. Our proposed method involves part of the electric vehicle charging technology, specifically involves the charging method of electric vehicles with multi battery series mode.

## IV. CONCLUSION

We propose a charging method for electric vehicles using multi battery series mode, which is characterized by the following steps: firstly, the battery series is charged at a constant power with a charging current of I1; when the terminal voltage over the battery series has reached the 1st threshold voltage, the charging current will reduce to I2 and the power remains constant; when the terminal voltage over the battery series has reached the 2nd threshold voltage, the charging current will reduce to I3 and the power remains constant (the 2nd threshold voltage is larger than the 1st threshold voltage); when the terminal voltage over the battery series has reached the 3rd threshold voltage, which is the rated voltage of the battery series, the charging process is completed (the 3rd threshold voltage is larger than the 2nd threshold voltage) The charging method is also characterized in the aspect that the charging controller operated in dynamic control mode is able to carry out real-time control of the input current and terminal voltage of the battery series. The terminal voltage over the battery series will gradually increase before it reaches the 3rd threshold voltage and the charge process is terminated by the charging controller.